\documentstyle[epsf]{elsart}
\begin{document}
\begin{frontmatter}
\title{The growth dynamics of German business firms}
\author{Johannes Voit\thanksref{elec}}
\address{Theoretische Physik 1, Universit\"{a}t Bayreuth, 
D-95440 Bayreuth (Germany), and Fakult\"{a}t f\"{u}r Physik, 
Albert-Ludwigs-Universit\"{a}t, D-79104 Freiburg (Germany) }
\thanks[elec]{e-mail: johannes.voit@uni-bayreuth.de, \\
web: http://www.phy.uni-bayreuth.de/$\tilde{~}$btp314}
\maketitle

\begin{abstract}
We determine the distribution of size and growthrates of German business firms
in 1987-1997. We find a log-normal size distribution. The distribution of 
growth rates has fat tails. It can be fitted to an exponential in a narrow 
central region and is dominated by finite-sample-size effects far in its
wings. We study the dependence of the growth rate distribution on firm size: 
depending on procedures, we find almost no
dependence when the center of the distribution is considered or, similar
to previous work,  a power-law
when the wings are weighted more strongly. Correlations in the growth
of different firms are essentially random. We determine the annual
growth of the entire economy, and successfully correlate it with a standard
economic indicator of business cycles in Germany. We emphasize possible 
problems related to the finite number of firms comprised in our database
and its short extension in time. 
\end{abstract}
\end{frontmatter}

\section{Motivation, important problems}

To understand the dynamics of economic growth, and its underlying 
mechanisms, is important for society. 
Important issues are, among others, the 
influence and control of economic growth
by government policy (e.g. taxes, employment conditions, or monetary policy);
industrial concentration and antitrust policy, and its influence on 
competition and economic growth;
changes in the structure of economies brought about by the globalization
of economic activities, and their consequences. Also, it is unclear if firms
which are part of the ``New Economy'' of high-technology sectors grow according
to the same rules as the firms of the ``Old Economy'', and if the same
macroeconomic indicators should be used, e.g., to measure the business
cycles of both economies. 

Surprisingly, it turns out, however, that the dynamics of economies is
not well understood, both empirically and with respect to its underlying 
mechanisms. \em ``Unfortunately, as is obvious from reading the newspapers,
the theory of macroeconomics is not a settled field. There is much 
controversy among economists what is a useful basic approach as well as 
about the detailed analyses of particular economic events and policy 
proposals'' \rm \cite{barro}. Economic growth,
or the concentration of firms in an industry are often measured by some
indicators, and the underpinning of these indicators by statistical data
on an economy sometimes remains unclear. 
Quite generally, economic theories often focus on mean values, trends, 
macrovariables \cite{barro,bp}.

Firm sizes in classical theory would be determined by 
the long-run cost curve:  
the cost of output has a U-shaped dependence on the amount of output,
and is minimal at a specific scale. Firms therefore would
grow or shrink under the influence of competition in the market, 
in order to attain this optimal size for maximal profits, and an equilibrium
would establish at that size. The question, of course, is if 
and to what extent, the distributions of firms in an industry, or an entire
economy, are consistent with such arguments and market mechanisms. 
It turns out, however, that both the arguments about, and even more so  
the evidences for, such ``economies of scale'' are
controversial \cite{ijisi}. 

A different approach is taken by
stochastic growth models \cite{ijisi} which 
attempt to describe the distribution of firm sizes in an economy, and
relate details of these distributions to elements of the growth dynamics.
The first such model was formulated by Gibrat \cite{gibrat} who postulated
that the relative rate of growth of a company is independent of its size.
Assuming further that the growth rates are uncorrelated in time and that firms
grow independently of each other, he arrived at a description in terms of 
a multiplicative stochastic process leading to a log-normal distribution of
firm sizes
which was verified by his observations. Later work by economists also
discussed other distribution functions involving power laws, but the evidence
for or against a specific class of distributions has remained controversial
\cite{ijisi}. Moreover, the discussion was strongly focussed on the shape
of the size distribution and a correlation of this distribution with
other properties of the sample was rarely attempted. 

In the past few years, physicists have become interested in 
of economic problems, mostly financial markets \cite{finmar}. It is 
tempting therefore to consider the dynamics of economic growth from a 
physicist's perspective.
Growth processes are routinely investigated and modelled in Statistical 
Physics \cite{growfor}. Examples are diffusion limited aggregation, or the 
Eden model, which may approximately describe the growth of crystals or tumors,
respectively. Here, the identification of a specific growth mechanism is based
on the form of the resulting object, i.e. on the correlation of its size
(volume or mass) with its local fluctuations. 
This brings up the question:
What can be learned from the study of fluctuations in economic growth? 

This question is not new, and some papers have adressed growth properties
of industries and economies \cite{amar,lee,taka,obert}. Amaral \em et al.,
\rm and Lee \em et al., \rm found that (i)
the size distribution of US firms and
of the gross domestic product of 152 countries approximately follow 
log-normal distributions; (ii) the distribution of growth rates conditioned
on the size of the firms (countries) are exponential; (iii) the standard
deviation of this exponetial distribution depends on size as $\sigma \sim
S^{-\beta}$ with $\beta \approx 0.16$ both for the US firms and for the 
countries. These findings were suggested to be universal, and indicative
of the growth dynamics of systems with complex internal structures
\cite{amar,lee}. Takayasu and Okuyama \cite{taka} performed a similar
analysis for different countries. They found that the shape of the firm
size distribution depended on the country (power-law for the US, more
exponential for France, Japan, and Italy) but obtained growth properties
consistent with (ii) and (iii) above. Finally, Ramsden and Kiss-Hayp\'{a}l
established the rank-size relation (Zipf analysis) for firms in the economies
of 20 countries in 1994 \cite{rams}. Unless there is clustering at  
certain sizes, this procedure
is rather equivalent to analyzing the cumulative distribution function.
They found significant differences among the countries in their fits. 

Interpretations were proposed in terms of systems with complex internal
structure but no competition between firms \cite{amar,lee}, competition
of structureless firms \cite{taka}, or thermodynamics \cite{rams}. The
simulation of these models could reproduce essential features of the 
empirical data in all cases. 

Important questions remain. One of the most immediate is about universality.
How universal are the statistical properties of economies? Another problem
is the correlation of a statistical analysis with economic indicators. 
To better understand these issues,
we perform an analysis of the growth properties of German business firms.

\section{Method of present work}
We investigate two database of German business firms: 
(i) Datastream provides a data base with 570 stock companies over the 11 
years 
1987-1997; (ii) the Hoppenstedt data base contains about 6500 firms over 
20 years. The investigation of this base
is not completed at the time of writing so that only results based on 
the more limited database
(i) will be presented. 

We use the annual sales as an indicator of the
size of a firm. Earlier work for US firms has shown, however, that other
indicators such as assets or the number of employees, give similar results
\cite{lana}. Due to the small sample size, we do not attempt to consider
different industries. Earlier work for other countries shows that the
results do not depend on the type of industry within a specific country
\cite{taka,lana}. Finally, we discard all firms with incomplete data from
our sample. The structure of the sample suggests that in most cases, data are
just ``missing'' randomly. However, we eliminate in this way also effects
of mergers and acquisitions, as well as 
newly founded firms and firms going bankrupt. 

Our data base finally contains
405 companies with sales data for the 11 years 1987-1997. Their sizes will
be denoted by $S_j(t_i), \; j=1,\ldots,405, \; i=1987,\ldots,1997$.
One source 
of concern is the small size of this sample. Is it representative for 
the German economy? We attempt a preliminary answer by comparing to a
standard indicator of economic growth below. More definite statements, 
however, will have to be based on  studies using the bigger
data base (ii), to be published elsewhere. Apart the limited number of
companies sampled, 
we will point out repeatedly the problems associated with the finite extent
in time, of our data base, and the limitations they imply for interpretation.
Notice that many other studies suffer from similar limitations. 

\section{Firm size distributions, annual growth}
\begin{figure}[b!]
\epsfxsize=12.5cm
\centerline{\epsfbox{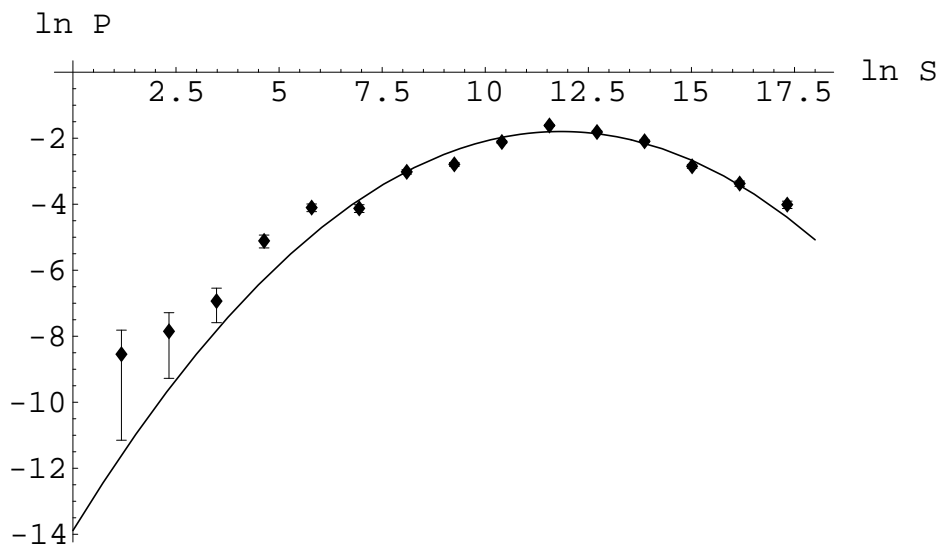}}
\caption{Size distribution of German business firms. Dots: empirical data,
line: fit to a log-normal distribution, parameters 
$\langle \ln(S) \rangle = 11.8, \; \exp \langle 
\ln(S) \rangle =  137 \times 10^6$ DM, standard deviation = 2.4.  
}
\label{fsdist}
\end{figure}

Figure 
\ref{fsdist}
shows the firm size distribution of the entire 
database $P[S_j(t_i)]$, i.e. all years 1987-1997 mixed. 
Similar results are obtained, 
though with less good statistics, when individual years are analyzed. 
The size distribution of German business 
firms indeed is approximately log-normal, in agreement with Gibrat's 
observations \cite{gibrat} and the US firms \cite{amar}. 
We note, however, that
there is a sharp cutoff at big company sizes, and an excess of weight
on the small-size wing of the distribution. These features are observed
systematically with a one-year resolution, too. The cutoff at big sizes
is consistent with similar observations for Japan, France, and Italy
\cite{taka} but the data presentation in that work does not allow to draw
conclusions on the small-size limit. We have not checked fits to alternative 
distribution functions.

\begin{figure}[b!]
\epsfxsize=12.5cm
\centerline{\epsfbox{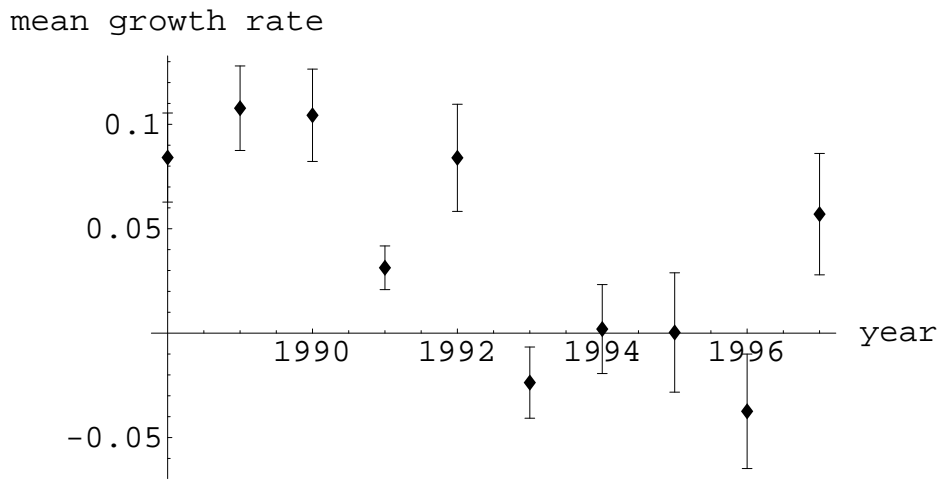}}
\caption{Mean growth rate of the German economy as sampled by our database
in 1987-1997.}
\label{anngrem}
\end{figure}

From the firm sizes $S_j(t_i)$, annual growth rates $r_j(t_i)$ are derived as
$r_j(t_i) = \ln \left[ S_j(t_i) / S_j(t_{i-1}) \right]$.
The mean of the distribution of the annual growth rates gives the growth
of the entire economy (to the extent that our limited database gives a 
faithful representation), and is shown in Figure \ref{anngrem}. 
Despite the scatter, one can notice an apparently systematic variation of 
the economic conditions in Germany: there is a boom period in the late 80's
and early 90's, slowing down and even recessing during the mid 90's and 
a restart of positive growth at the end of the sampling period of our 
database. While for residents of Germany this pattern will remind, and
correlate with, the media reports on the economic conditions of the country,
it is necessary to compare with established indicators of business cycles
and economic trends,
in order to assess the relevance of our analysis.

\begin{figure}[t!]
\epsfxsize=12.5cm
\centerline{\epsfbox{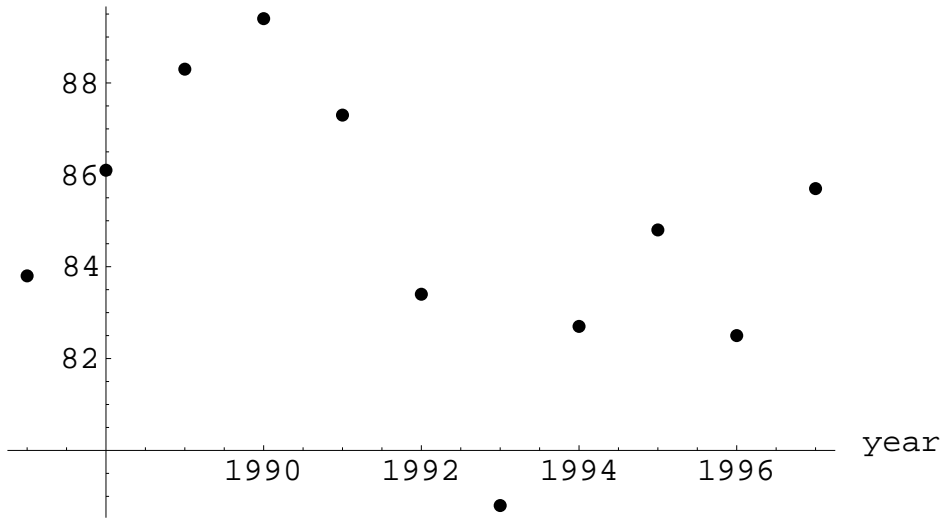}}
\caption{Percentage use of production capacity of German manufactoring sector
in 1987-1997.}
\label{jagu1}
\end{figure}

One such indicator for business cycles
is provided by the (percentage) use of production capacity
of German business firms \cite{bp}. 
Data for the manufactoring sector are reported 
annually by the Advisory Panel to the German Federal Government on 
the Economic Conditions, and shown
in Figure \ref{jagu1} 
for the same period as our database. They show a pattern similar to
the annual growth rate derived from our statistical analysis. 
This suggests that our data base may indeed provide a good representation
of the German economy, despite its limited size; 
conversely, the agreement shows that the use of such coarse-grained 
indicators can be backed by a more ``microscopic'' statistical analysis.
A priori, it is not obvious if this quantity, or rather its annual change
should be compared to the mean growth rates shown in Figure \ref{anngrem}.
Its annual change, however, correlates less well with our analysis. We 
therefore conclude that the use of production
capacity itself is the appropriate indicator. 

\section{Growth rate distributions}
\begin{figure}[b!]
\epsfxsize=12.5cm
\centerline{\epsfbox{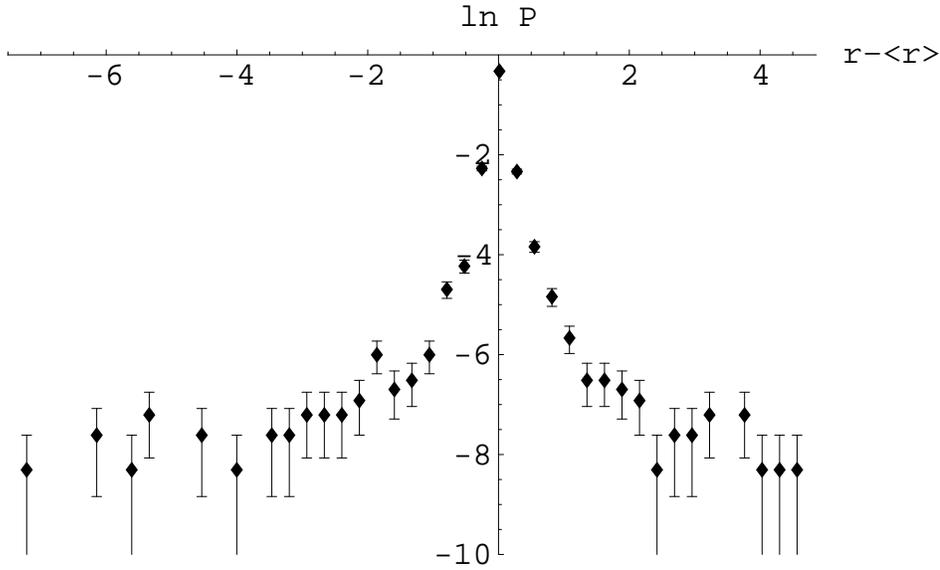}}
\caption{Distribution of growth rates of German business firms
with respect to their mean in 
1987-1997.}
\label{distgr}
\end{figure}

We now turn to the distributions of the firm growth rates. Figure \ref{distgr}
shows the distribution of the fluctuations in the growth rates (i.e. after
the subtraction of the mean $\langle r_j(t_i) \rangle_{i,j}$), 
for the entire database, all years mixed. 
Apart the
less good statistics, however, the distributions time-resolved on a one-year
scale, show the same behavior. It is clear that the raw data have a 
fat-tailed distribution, far from both the normal distribution associated
with a simple multiplicative stochastic process, and from the exponential
distribution which has been found in earlier work \cite{amar,lee,taka,obert}. 
At present, we have not attempted to fit the entire distribution to a 
particular form. We rather analyze separately the center and the wings of
the distribution. 

\begin{figure}[b!]
\epsfxsize=12.5cm
\centerline{\epsfbox{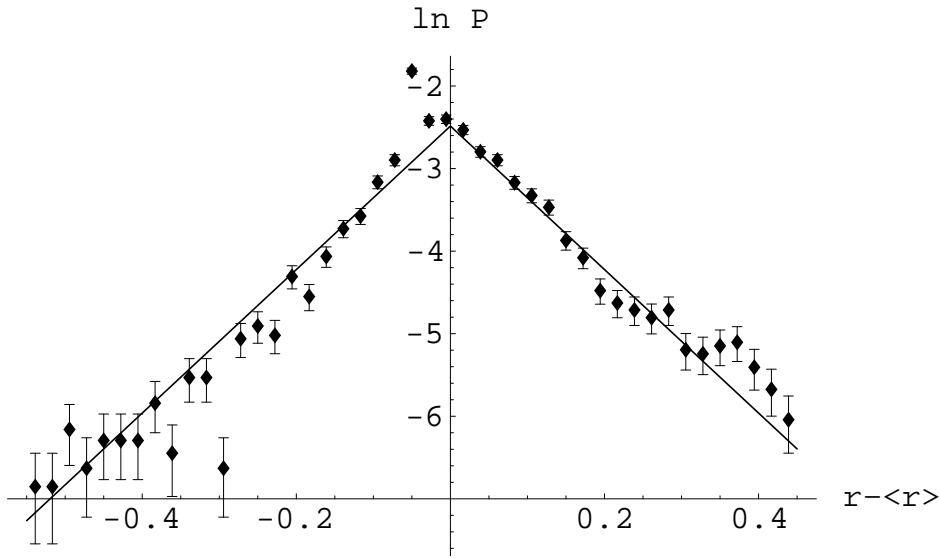}}
\caption{Central part of growth rate distribution of German business firms
in 1987-1997, and fit to an exponential distribution (solid line). 
3788 out of 4050 data
have been used in this analysis.}
\label{grcent}
\end{figure}

\begin{figure}[t!]
\epsfxsize=12.5cm
\centerline{\epsfbox{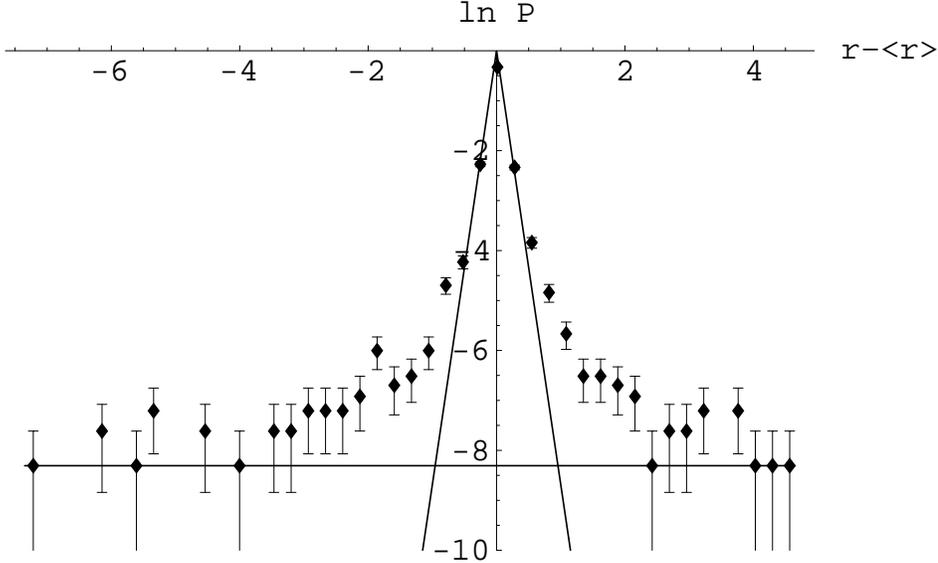}}
\caption{Superposition of growth rate distribution of German business firms
with central fit to an exponential distribution (tent-shaped solid line) and
lower-finite size cutoff. The horizontal line corresponds to one count per 
bin.}
\label{synth}
\end{figure}

Figure \ref{grcent} shows the distribution of the growthrates of firms with
$-0.55 \leq r_j(t_i) - \langle r \rangle   \leq 0.45$.
In this range, a successful fit to an exponential distribution is possible 
indeed (standard deviation 0.12). Hence we recover the exponential growth rate
distributions found by Amaral \em et al., \rm \cite{amar} and Takayasu and
Okuyama \cite{taka}, albeit only for a very narrow range of growth rates.
The wings of the probability distributions are dominated by effects of 
finite sample size. The horizontal line in Figure \ref{synth}, where the dots
are the data shown in Figure \ref{distgr}, represents the
lower limit on counting, one count per bin. 
In the center of the figure,
the exponential fit function derived in Figure \ref{grcent} is superposed
on the original data. Again, all features are recovered in time-resolved data
on a one-year scale, with less good statistics.

Some observations, however, make us hesitate to endorse an exponential
growth rate distribution at the present level of data analysis without
reservations. They are related to
the extremely reduced range over which the exponential fit can be
performed.  Counting rates are still 
rather high (50-200 firms per bin) in the range where deviations from 
exponential behavior are clearly marked already, and the deviation of the
data from the exponential fit in the center is much larger than the error bars.
The parameters
derived from an exponential fit depend on the range of data used, and do 
\em not \rm converge well as this range is reduced. Finally, the shape
of our bare growth rate distribution, Figure \ref{distgr}, 
is very different from those found in other work \cite{amar,taka}, and more
reminiscent of work on stock exchange crashes which have been claimed to be
outliers with respect to a presumed exponential distribution \cite{sorjo}.

\begin{figure}[b!]
\epsfxsize=12.5cm
\centerline{\epsfbox{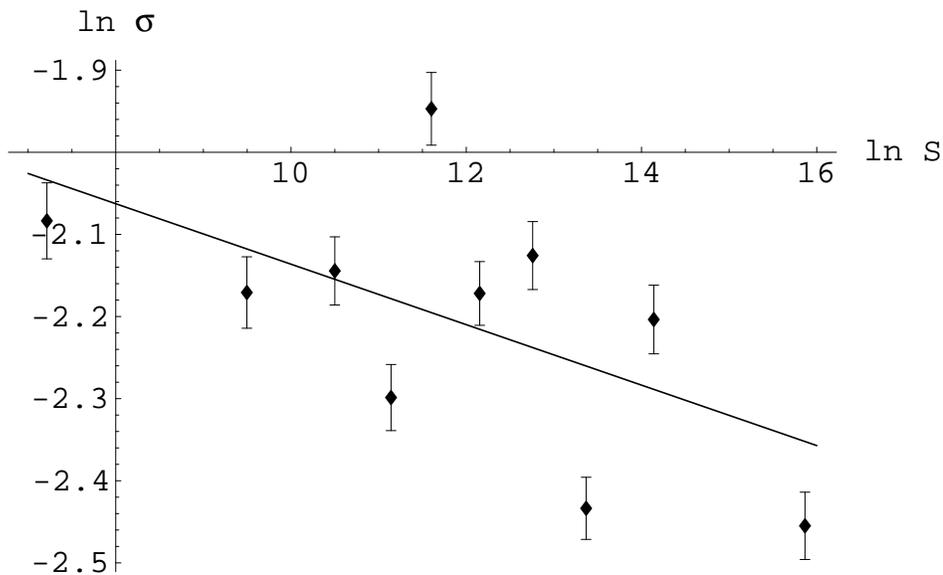}}
\caption{Dependence of the width $\sigma$ of the exponential distribution
fitting the center of $P(r- \langle r \rangle)$ vs. firm size. No clear
dependence emerges. The solid line is the ``best'' fit to a power law 
$\sigma \sim S^{-\beta}$ with 
an exponent $\beta = 0.037$.}
\label{wisi}
\end{figure}

\begin{figure}[b!]
\epsfxsize=12.5cm
\centerline{\epsfbox{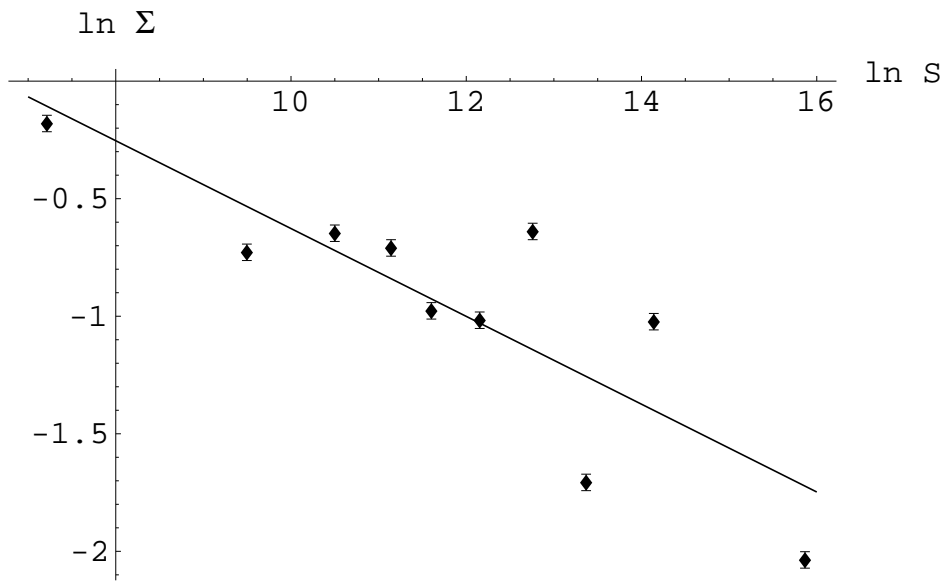}}
\caption{Dependence of the 
standard deviation $\ln \Sigma$ of the individual bins
on firm size. The solid line is a fit to a power law $\Sigma \sim S^{-\beta}$
with $\beta = 0.19$. }
\label{wist}
\end{figure}

Earlier work 
made the interesting observation that both for firms, and 
for the gross domestic product of countries, the
width $\sigma$
of the exponential growth rate distribution depends on the sizes of the firms,
resp.\ countries, when the data set is binned according to firm or country
size \cite{amar,lee,taka}. 
A power law dependence $\sigma \sim S^{- \beta}$ was found with a 
rather universal exponent $\beta \approx 0.16$. 
Figure \ref{wisi} shows the corresponding result for German business firms.
To generate this figure, we divided our firms into ten bins corresponding to
the $n/10$-quantiles of the sample. We then determined,
the distribution of growth rates for each bin, and performed
a fit to the center of the distributions as explained above. The resulting
widths $\sigma$ are plotted vs. the average firm size in the respective bin,
in Figure \ref{wisi}.
There is a large scatter of points in
the $\sigma$-$S$ plane, and no clear dependence of width on size emerges.
If we -- reluctantly -- attempt a power law fit, we find an exponent 
$\beta = 0.037 \pm 0.006$, essentially zero. If, instead, we use the standard
deviation $\Sigma$ of 
the entire sample in each bin, we do recover the power law $\Sigma \sim
S^{-\beta}$ with an exponent $\beta = 0.19 \pm 0.02$, Figure \ref{wist}.
This power law
dependence essentially agrees with the earlier results \cite{amar,lee,taka},
and supports the claims of universality made there.
Notice, however, that this analysis more strongly
reflects the variations in the wings of the growth rate distribution which
may deviate from the exponential statistics discussed above, and which may
suffer from severe finite sample-size effects. 
In the future, it will be important to understand the consequences of the
dependence of the exponent $\beta$ on the sampling procedure.

It will also be important to study the reasons for the serious
discrepancies between fitted laws and empirical data, in particular in 
Figures \ref{wisi} and \ref{wist}. In all figures, error bars are 
for one standard deviation. The number of data points deviating from the 
fitted laws in general is rather larger than 32\%. Similar conclusions would
be reached at the 95\% confidence level. It is not clear that going to larger
samples would improve the situation: quite generally, the error bars will 
decrease with sample size. Also, the apparently random scatter of the data
points around the fitted lines in Figs.\ \ref{wisi} and \ref{wist} leaves
little room for an improvement in the specification of the theory. However,
there is the possibility that one of the assumptions underlying the error
analysis is not satisfied: randomness of the data, i.e. growth rates. 
If the growth rates are strongly 
correlated, the number of independent entries in each of the
histogram bins will be reduced significantly, and the error bars will be 
correspondingly increased. 

\section{Correlations in firm growth}
Many models describing the observed firm size distributions, and those
describing the exponential growth rate distributions \cite{amar}, assume an
uncorrelated growth of firms both in time, and across the economy. 
This assumption can be checked, in principle,
from our data set. From an economic point of view, an assumption of 
uncorrelated firm growth may appear questionable, a priori. One the one
hand, when a national economy evolves significantly and non-randomly, 
this evolution should be visible in the growth dynamics of its members,
and is likely caused by correlations of the individual growth processes. 
In periods where people buy more (less) cars, the sales of Daimler-Chrysler,
Volkswagen, and BMW will increase (decrease)
while changes in market share of the 
individual companies operate on much smaller scales. On the other hand, 
correlations between car producers and, e.g., pharmaceutical companies
are less obvious and may even be negative. 

\begin{figure}[b!]
\epsfxsize=12.5cm
\centerline{\epsfbox{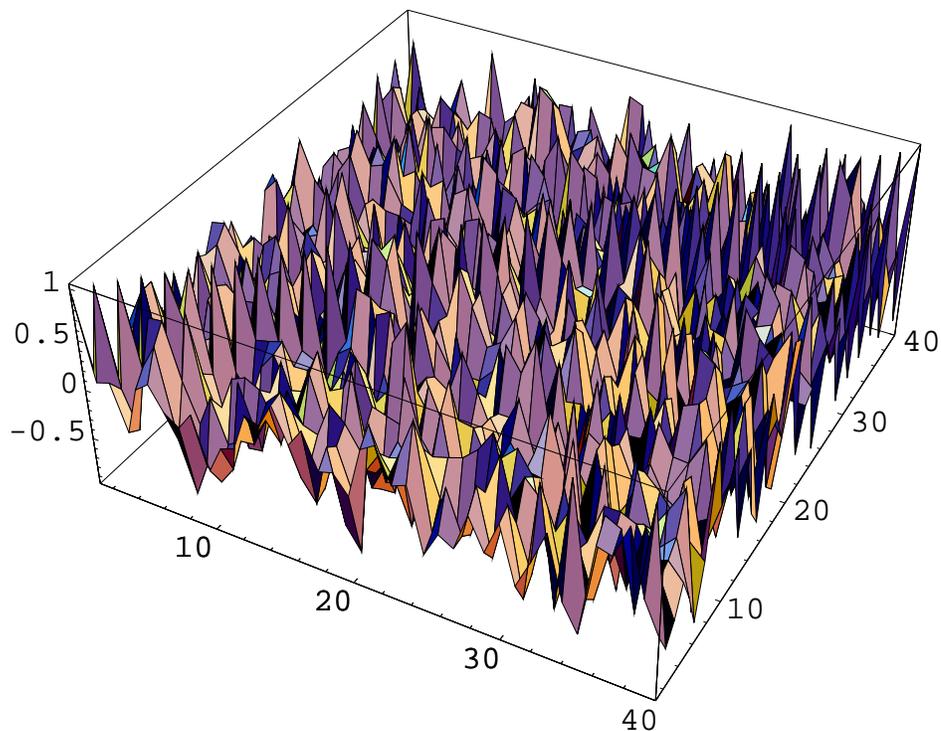}}
\caption{Covariance matrix 
$cov(i,j)$ of the growth rates of German business firms. For
clarity, we only show the first $40 \times 40$ entries.}
\label{cov1}
\end{figure}

\begin{figure}[t!]
\epsfxsize=12.5cm
\centerline{\epsfbox{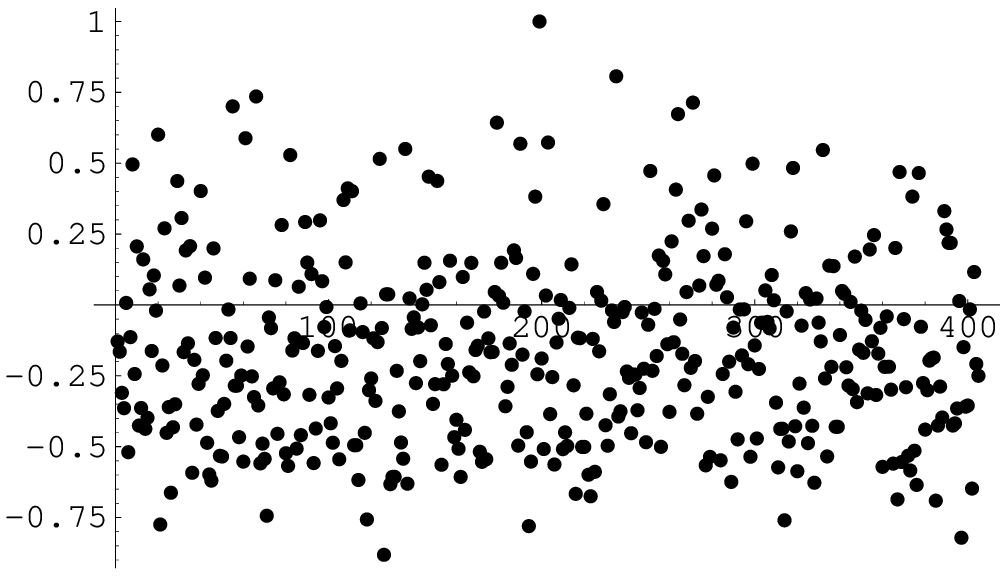}}
\caption{Cut through the covariance matrix, $cov(199,j)$, of German 
business firms.}
\label{cov2}
\end{figure}

We have computed the covariance matrix of the 405 firms of our database,
and show results in Figures \ref{cov1} and \ref{cov2}. 
For reasons of clarity,
Figure \ref{cov1} displays only the upper left $40 \times 40$ entries of 
the entire covariance matrix. Figure \ref{cov2} shows a cut through the
covariance matrix at row 199. Both figures apparently indicate that there
are significant correlations in the growth of the firms [uncorrelated 
growth would lead to $cov(S_i,S_j) = \delta_{i,j}$], and that these 
correlations are essentially random. Recent studies of financial markets 
indicate that correlations between stock prices also
have a strong random component
\cite{laloux}. There, about 94\% of the correlations are consistent with pure
randomness, and only 6\% deviate significantly from such a hypothesis. 
One might then ask: Are economies glassy systems?

\begin{figure}[b!]
\epsfxsize=12.5cm
\centerline{\epsfbox{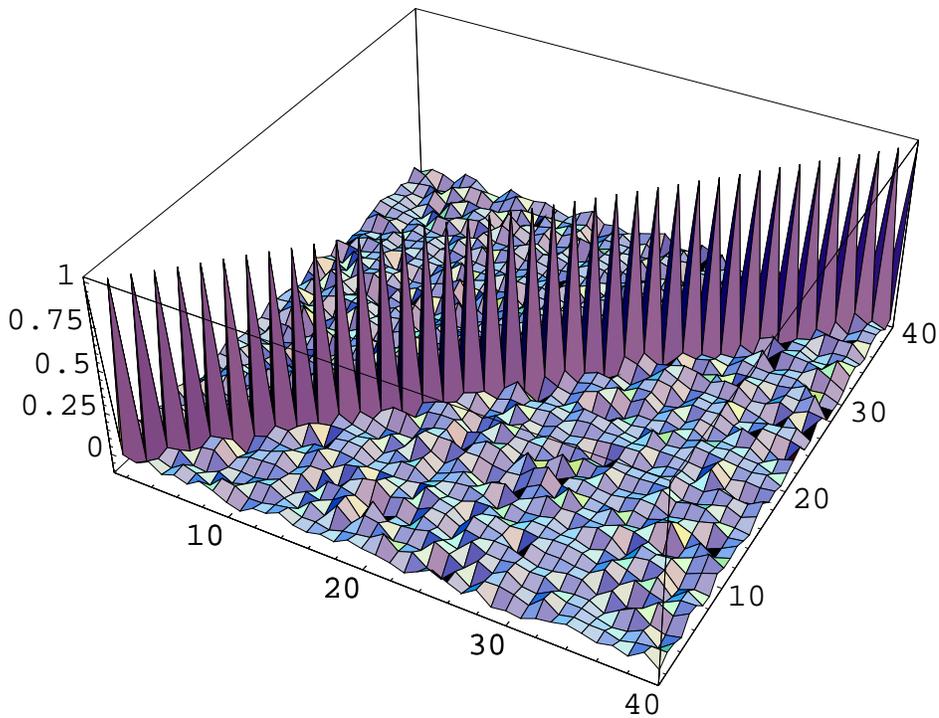}}
\caption{Covariance of 405 series of random numbers with 1000 elements in
each series. Only when the length of the series is comparable to the number
of series, does the uncorrelated nature of the random numbers emerge.}
\label{covran}
\end{figure}

While the idea of random correlations in firm growth dynamics may have some
appeal (cf. above), we caution against premature conclusions in this 
direction. In fact, the apparent randomness, or at least part of it, may
be due to the small temporal extension of our data set. Sufficiently
long time series are needed to make significant 
cross-correlations between various
variables emerge, and clearly distinguish them from uncorrelated random 
variables. Even a rather large number of uncorrelated
random variables will show some correlations provided one does not look at 
too long a time series (``noise dressing''). This is demonstrated on a 
surrogate data set in Figure \ref{covran}.
In this data set, ``time series'' for 405 ``firms'' and lengths of 10, 100,
and 1000 entries (``years'') have been produced with a random number 
generator. While the covariance matrix for 10 ``years'' is essentially
indistinguishable from Figs.\ \ref{cov1} and \ref{cov2}, and correlations
are still pretty strong for a 100 ``years'', it is only for 1000 ``year'' long
time series that the statistical independence of the random numbers becomes
apparent. The upper left $40 \times 40$ entries of the covariance matrix for
this latter case are shown in Figure \ref{covran}.

\section{Discussion, Open questions}
The preceding analysis of the growth dynamics of German business firms
shows notable similarities and differences to earlier work analyzing firms
in other countries. The list of similarities include a 
log-normal size distribution, and an exponential distribution of growth rates
at least in the center of the distribution, albeit with some reservations
with respect to the entire distribution.
Differences are found (i) in certain systematic deviations from the log-normal
distribution which is very well observed by US-firm or country data 
\cite{amar,lee}, and which may  reflect specific features of the German
economy, and (ii) 
in the dependence of the width of the growth rate
distribution on firm sizes where no clear dependence emerges when the
width of the exponential distribution in the center is used. However,
we do observe a power law $\sigma \sim S^{-\beta}$ as found elsewhere
\cite{amar,lee,taka} when the standard deviation of the entire
sample in each bin is evaluated. It is not clear if the difference in the 
exponents ($\beta = 0.19 \pm 0.02$ here vs.\ $0.16$ elsewhere) is significant.

The cutoff in the firm size distribution for big firms could be related
either to the size of Germany (the country can't support firms beyond a
certain size) or, in a more global perspective, 
to the fact that the biggest multinational companies are based in the US.
The excess of weight for small company sizes is much less clear and will
require further investigation. It may possibly be related to a traditionally
big number of small and medium-size enterprises in Germany 
(``Mittelst\"{a}ndische Wirtschaft'') which are well supported by politics, 
if they comprise a sufficient number of stock companies and therefore show
up in our sample.

Several models have attempted to describe the size and growth rate 
distributions and the dependence of the latter on firm size, based
on a distribution of minimal firm sizes necessary for survival in the
different industries composing an economy \cite{amar}, on the amount of
competition in an economy and its control through, e.g., taxes \cite{taka},
or on managerial culture (lean management vs. hierarchical and authoritarian)
\cite{buldy}. It will be interesting to interpret the size-independence of the
growth rate distribution
found here for the German economy, with specific
features of these models, and to correlate these with independent economic
information on German business firms. This is planned for future work.
If successful, such a program could open
the exciting possibility to correlate statistical properties of an
economy with political, historical and cultural elements, and to closely
monitor the changes brought about by both European integration and the 
worldwide globalization of the economy.

\begin{figure}[t]
\epsfxsize=12.5cm
\centerline{\epsfbox{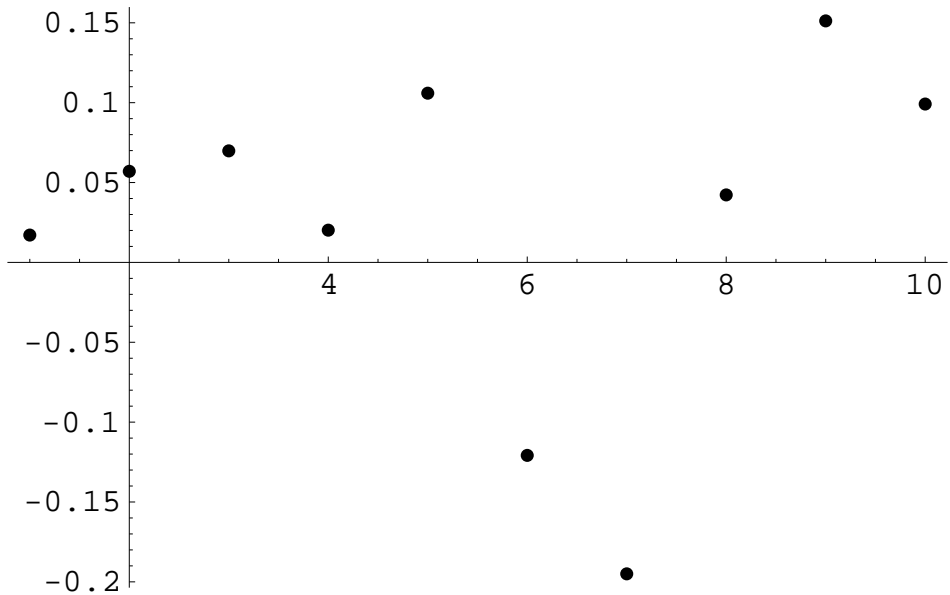}}
\caption{``Annual growth'' of a surrogate economy (random numbers). 
Systematic and random variations cannot be distinguished for few entries 
only.}
\label{ragro}
\end{figure}

At the interface between statistics and economy, we have shown that the
temporal evolution of the firms listed in our database tracks standard 
indicators of economic growth. On the other hand, we again caution against
premature conclusions: Figure \ref{ragro} shows the average growth of a
surrogate economy (random numbers), and the differences to Figures 
\ref{anngrem} and \ref{jagu1} apparently are minor. 
The similarity of these figures makes us wonder to what extent the variations
in the growth of the German economy are systematic, i.e. reveal business 
cycles, and how big a random component they contain. On what time scales 
does non-randomness become important? Data for the growth rate of the gross
national product of the US 1870-1988 look random superposed on a small 
positive offset while those for Germany 1950-1992
show a more periodic structure \cite{grat}. 
This points to an
important limitation of economic time series, compared to time series 
in physics, finance, medicine, etc., namely their short 
extension in time. Gross indicators are available over rather long times
but detailed firm data have been collected rather recently only.
Finite sample length
effects are a serious problem, and may also have affected the analysis 
presented here. 

Future work therefore should address temporal correlations, and the
importance of business cycles. Some evidence
in favor of temporal correlations has been gathered in the past \cite{ijisi}. 
It would also be of interest to investigate birth and death processes in
the economy. 
Finally, it will be interesting to confront the rather
pragmatic approach of a physicist used here, to the more elaborate 
tools used by econometricians such as hypothesis tests, etc. 
These issues will be taken up in future publications. 

\section{Acknowledgements}
I am a Heisenberg fellow of Deutsche Forschungsgemeinschaft. Most of
all, I would like
to acknowledge advise from and discussions with  Dirk Obert (Freiburg) who
also provided the data used here. This contact has been instrumental in
getting this work started. I also would like to acknowledge discussions with
S. H\"{u}gle, C. Jelitto, C. Kreuter, V. Plerou, and H. E. Stanley.

\end{document}